\documentclass[aps,preprint,onecolumn,eqsecnum,nofootinbib]{revtex4}

\usepackage{epsfig}
\usepackage{graphicx}
\usepackage{dcolumn}
\usepackage{amsmath}
\usepackage{enumerate}
\usepackage{subfigure}

\begin{document}
\vskip-6pt \hfill {IPPP/09/06}\\
\vskip-6pt \hfill {DCPT/09/12}\\

\title{On the possibility of sourcing a mono-energetic $\bar{\nu}_{e}$ long baseline beta beam from bound beta decay}
\author{Christopher Orme}
\affiliation{IPPP, Department of Physics, Durham University, Durham
  DH1 3LE, United Kingdom}

\begin{abstract}
In this paper, the possibility of using fully stripped ions that can decay through bound beta decay to complement electron
capture long baseline neutrino oscillation experiments is qualitatively analysed. The disadvantages of such a source are 
discussed through consideration of the technological challenges faced and the energy 
resolution required from the detector. It is concluded that ions that bound beta decay cannot be used as a source 
of mono-energetic anti-neutrinos in a realistic long baseline CP-even neutrino beam.

\end{abstract}

\maketitle

\section{Introduction}

Bound beta decay is the process in which, instead of ejecting an electron, as in usual beta decay, the electron becomes
bound to the daughter ion. The anti-neutrino energy spectrum is therefore not continuous, instead 
it is a series of discrete mono-energetic spectra with the intensities dependent on the electron orbital occupied. 
Bound beta decay (BBD) was predicted in 1947 by 
Daudel \emph{et al.} \cite{rD47} and was discussed theoretically by Bachall \cite{jB61}. 
However, it was not until 1992 that BBD was confirmed experimentally in $^{163}$Dy$^{66+}$ at the GSI Helmholtz
Centre for Heavy Ion Research \cite{firstobs}. 

In the last couple of years, the BBD process has been considered for very short baseline neutrino oscillation experiments with very small detectors also; 
$L\sim 10$ m and detector mass $\sim 100$ g~\cite{moss}. Following works by Raghaven~\cite{raghaven}, it is proposed to 
use the processes
\begin{equation}
^{3}\rm{H} \:\:\longrightarrow \:\:^{3}\rm{He} + e^{-} + \bar{\nu}_{e}\qquad \mbox{and} \qquad
^{3}\rm{He} + e^{-} + \bar{\nu}_{e} \:\:\longrightarrow \:\:^{3}\rm{H}   
\end{equation}
as the emission and detector processes, respectively. The electron in each case is bound. By embedding the $^{3}$H and 
$^{3}$He atoms in a metal lattice, the detection process becomes resonantly enhanced with cross-sections up to 12 orders of
magnitude larger than a non-resonant capture at the same energy. Oscillation of these `Mossbauer neutrinos' has 
been demonstrated theoretically~\cite{mossosci}, and they have been studied in the context of determining the neutrino 
mass hierarchy 
without using matter effects~\cite{moss} and in the search of active-sterile neutrino oscillations, amongst other things. 
In this
paper, however, only long baseline experiments will be considered and so this possible use of the BBD process will not be
discussed further. 
The use of the BBD process in long baseline neutrino oscillation experiments has recently been suggested~\cite{CPeven,Maurotalk}, either for use in its own right or in view of complementing electron capture machine proposals~\cite{catalina,sato,satorolinec}. 
In this short article some practical problems with this proposal are identified and qualitatively 
analysed. The very low branching ratios of the BBD process; large proton number of the required ions, and the energy resolution of the far detector all indicate
that such a machine is not practical.


\section{Bound beta decays}\label{S:BBD}

A continuum beta decay (CBD) is a transition between discrete stationary states of a parent and daughter nucleus - 
usually from ground state to ground state or low lying 
excited state. The $\beta^{-}$ CBD process transforms a neutron bound within a nucleus into a proton accompanied by the 
creation of an electron and anti-neutrino:
\begin{equation}
^{A}_{Z}X \longrightarrow \; ^{A}_{Z+1}Y + e^{-} + \bar{\nu}_{e}.
\end{equation}
The decay electron is virtually always emitted to the continuum with capture in the outer orbitals strongly suppressed due
to the weak bindings and small wave function 
overlaps. Capture into the inner orbitals is clearly forbidden by the Pauli principle. The electrons from $\beta^{-}$ for 
a fully ionised parent $^{A}_{Z}X^{Z+}$, however, can be 
captured into a bound orbit: 
\begin{equation}
^{A}_{Z}X^{Z+} \longrightarrow \; ^{A}_{Z+1}Y^{Z+} + \bar{\nu}_{e},
\end{equation}
with the $\bar{\nu}_{e}$ being mono-energetic. 


The kinematics for neutral and fully ionised atoms are not the same; 
corrections are necessary to compensate for electron binding energies. The CBD Q-value for ion with atomic mass $A$ and 
proton number $Z$ is defined as
\begin{equation}
Q_{C} = m_{Y}(A,Z+1) - m_{X}(A,Z),
\end{equation}
where the mass of the ejected electron has been implicitly included. The $Q_{C}$ is therefore the kinetic energy 
available to the channel, for an atom with the full complement of orbital electrons. When the atom is stripped bare, the 
Q-value needs to be corrected by the difference in binding energies of the complete parent and daughter ions,
$\vert \Delta B^{tot}_{Y,X} \vert$. To get the BBD process Q-value, $Q_{B}^{Z+}$, this needs to be further modified by the 
binding energy of the electron captured into orbital $n$ of the daughter nucleus, $\vert B_{n;Y}\vert$. 
In summary~\cite{dB05}, we have:
\begin{equation}
\begin{array}{rll}
Q_{B}^{Z+}&= Q_{C}&+\:\vert B_{n;Y}\vert -\vert \Delta B^{tot}_{Y,X} \vert \\
          &= Q_{C}^{Z+} &+\: \vert B_{n;Y}\vert.\label{E:binding}
\end{array}
\end{equation}
BBD is an important process in astrophysics owing to the heavily ionised environments. These corrections have therefore 
been tabulated for many ions~\cite{jP73,wJ85}. $\vert \Delta B^{tot}_{Y,X} \vert$ is the smaller correction to the 
$Q_{C}$, 17 keV for Thallium~\cite{jP73}, for example. $\vert B_{n;Y}\vert$ is the larger but still small compared to 
$Q_{C}$: 99 keV for Thallium \cite{wJ85}.

BBD can be thought of as the inverse process of electron capture. The relationship between the decay rates for BBD and 
CBD is therefore similar to that of CBD and electron capture. Taking BBD and CBD to have the same nuclear matrix elements
and using phase space arguments, for respective branching rates $\Gamma_{C}$ and $\Gamma_{B}$, the relative strength is 
given by~\cite{jB61}

\begin{equation}\label{E:ratio}
\frac{\Gamma_{B}}{\Gamma_{C}}\propto\frac{Q_{B}^{2}\:\vert \psi_{n}(0) \vert^{2}}{f(Z,Q_{C})},
\end{equation}
where
\begin{equation}
f(Z,Q_{C}) = \int_{m_{e}}^{Q_{C}+m_{e}}E\sqrt{E^{2}-m_{e}^{2}}\:(Q_{C}-E)^{2}F(Z,E)dE \label{E:f}
\end{equation}
is the integral over phase space for CBD and $\psi_{n}(0)$ is the wave function for an electron in the nth 
orbital.
It is seen that the strong $Q_{C}$ dependence for CBD means that this channel will 
not only dominate for high $Q_{C}$ but will have a substantial branching ratio at all but the very smallest $Q_{C}$. Since
the rate for BBD is dependent on the modulus square of the orbital wave function, there is a proton number dependence 
which leads
to substantial branching ratios for high Z atoms, provided $Q_{C}$ is not too high. A more detailed calculation has been 
carried out in~\cite{jB61} and the branching ratios, in a full relativistic calculation, have been presented 
in~\cite{dB05}. The ratios presented in Tab.~1 were calculated using the expression for $\Gamma_{\rm BBD}/\Gamma_{\rm CBD}$ given in~\cite{table}
without including radiative corrections.



\section{Technological challenges}\label{S:Disadvantages}

In this section, the possibility of using BBD will be discussed and some of the technological challenges articulated. In 
particular, the demands on the acceleration chain and the impact on the anti-neutrino fluxes will be first focussed on. 
In the second part, the effect of the energy resolution of the detector will examined with the likely demands on the
acceleration chain investigated. Although a concrete long baseline setup is not being proposed here, some ion 
`choices' are presented in Tab.~\ref{T:ions} to make the discussion more explicit. An optimal ion will have a half-life 
$\sim 1$ second~\cite{Autin:2002ms}; however, the paucity of choice means the half-lives may be much higher. A scan of the
database~\cite{ions} for a selection of ions with single dominant decay channels and half-lives in the range 
$0.5 \:\:\mbox{sec} < t_{1/2} < 8 \:\:\mbox{min}$ was made. Very few ions matched the criteria.    

\begin{table}
\begin{center}
\begin{tabular}{c|c|c|c|c}
Ion & Q-value & Channel \% & Half-life & $\Gamma_{BBD}/\Gamma_{CBD}$ \\
& (MeV) &&&\\
\hline
\hline
&&&&\\
$^{20}$O & 2.757 & 99.97 & 13.51 sec & $9.4\cdot 10^{-5}$\\
$^{34}$Si & 2.993 & 100 & 2.77 sec & $3.6\cdot 10^{-4}$\\
$^{52}$Ti & 1.831 & 100 & 1.7 min & $8.8\cdot 10^{-3}$\\
$^{56}$Cr & 1.506 & 100 & 5.94 min & $7.0\cdot 10^{-3}$\\
$^{55}$Cr & 2.603 & 99.96 & 3.497 min & $2.1\cdot 10^{-3}$\\
$^{62}$Fe & 2.023 & 100 & 68 sec & $4.5\cdot 10^{-3}$\\
$^{98}$Zr & 2.250 & 100 & 30.7 sec & $0.010$\\
$^{99}$Nb & 3.403 & 100 & 15.0 sec & $4.1\cdot 10^{-3}$ \\
$^{120}$Cd & 1.760 & 100 & 50.8 sec & $0.026$\\
$^{121}$In & 2.434 & 100 & 23.1 sec & $0.014$\\
$^{206}$Tl & 1.533 & 99 & 4.199 min & $0.080$ \\
$^{207}$Tl & 1.423 & 99.72 & 4.77 min & $0.138$\\
$^{209}$Tl & 1.832 & 98.8 & 2.20 min & $0.118$\\
\end{tabular}
\end{center}
\caption{A selection of ions selected based on their half-lives and dominant decay channels. The quoted Q-values are for
CBD and need to modified as discussed in Sec.~\ref{S:BBD} fully stripped ions.}\label{T:ions}
\end{table} 

\subsection*{Acceleration and flux}

In a beta beam, the radioactive ions are accelerated then stored in a ring to decay. To source a useful flux from 
the storage rings requires an optimal half-life 
$\mathcal{O}$(1 sec). The half-life needs to be sufficiently long to minimise losses in the acceleration, but
sufficiently short to source a useful flux once in a decay ring. This is one of the primary reasons why $^{18}$Ne, $^{8}$B,
$^{6}$He and $^{8}$Li are excellant candidate ions. The ions put forward for electron capture machines and BBD machines 
are not optimal in that they have half-lives up to several minutes and so the number of useful neutinos sourced is
several orders to low~\cite{Fraser}. This problem could be dealt with R\&D 
in the acceleration stage: increased production rates, reduction of losses during acceleration, and loosening of 
constraints on the duty factor could all lead to a boost in useful decay rate. An accumalation ring is also
an option to compensate for the accelerator complex dead time of approximately 8 seconds~\cite{Fut_prod}.
For electron capture machines, the aim is to choose ions with near 100\% branching ratios. This is not a luxury 
available to BBD sources however. 

The branching ratio for BBD is 
typically small
unless the Q-value is very small or the proton number of the ion is large. However, if one wishes to source a long 
baseline experiment, very small Q-value ions are not an option (Tab.~\ref{T:ions}). Low or modest
branching ratios are therefore an intrinsic feature of BBD long baseline candidate ions. Achieving the necessary count
rates is therefore very demanding; for example, consider $^{207}$Tl which has the highest branching ratio of the 
selected ions in Tab.~\ref{T:ions}. $10^{18}$ useful decays is the target rate for any long baseline beta beam type experiment. If this could be
achieved, one is still an order of magnitude short for the useful mono-energetic anti-neutrinos. In addition, to extract 
a useful BBD rate requires hydrogen-like atoms. A large proton number is likely which points to severe 
space charge issues, especially in the low energy part of the accelerator chain. These effects 
collectively force 
the need for an extra factor of 10 in 
production~\cite{lindroos} requiring an extensive R\&D program and large duty factors (up to 10\%). For a fully stripped ion, vacuum losses are not
a concern since the probability of the ion capturing an electron is effectively nil~\cite{Fut_prod}. 

The ions considered in~\cite{CPeven} could BBD, CBD and decay through electron capture. Four ions were identified with BBD
Q-values ranging from 1.67 MeV to 2.46 MeV. The branching ratios were therefore low ($\sim 1~\%$). The motivation behind this proposal 
was
to use the BBD and electron capture spectra with the end part of the CBD spectrum to construct a `CP-even' beam defined by
\begin{equation}
\eta(E;\gamma) = \frac{\mathcal{F}(\nu_{e})\sigma(\nu_{\mu})-\mathcal{F}(\bar{\nu}_{e})\sigma(\bar{\nu}_{\mu})}{\mathcal{F}
                                  (\nu_{e})\sigma(\nu_{\mu})+\mathcal{F}(\bar{\nu}_{e})\sigma(\bar{\nu}_{\mu})} =0~,
\end{equation}
where $\mathcal{F}$ is an unoscillated neutrino flux and $\sigma$ is a cross-section.
Such a strategy requires the separation of the neutrino and anti-neutrino events at the detector in addition to the separation of BBD and CBD events. One 
therefore needs to consider the characterisitics of the detector.


\subsection*{Detectors and energy resolution}\label{S:Dis} 

In the previous section, a number of issues surrounding the production and acceleration were highlighted. BBD will now be 
examined 
in the 
context of the likely technology available to the beta beam class of machines and what energy resolutions are 
required. 

 For an ion boost $\gamma$, 
an energy $E_{l}$ in the laboratory frame is related to its rest frame counterpart by $E_{l}=2\gamma E_{r}$.
For a given accelerator,  the maximum boost possible for an ion $_{Z}^{A}X^{N+}$ is given by
\begin{equation}
\gamma_{\rm{ion}}^{\rm{max}}=\frac{N}{A}\gamma_{p}^{\rm{max}},
\end{equation}
where $\gamma_{p}^{\rm{max}}$ is the maximum boost of the proton and N is the number of electrons removed from the atom.
 For the 1 TeV  machines available to beta beams, such as an upgraded Super Proton Synchrotron (SPS), 
$\gamma_{p}^{\rm{max}}=1066$. Ions that 
beta decay lie on the neutron-rich side of the line of stability on a 
Segre chart, and typically have $Z/A \sim 0.4-0.5$. Therefore, energies $\sim 1$ MeV in the rest frame correspond to 
energies $\sim 0.8$ GeV in the laboratory frame at maximum boost. In what follows, the lower limit, $\gamma_{\rm{ion}}^{max}=400$ is taken.

A beam source from ions that electron capture decay and bound beta decay will
contain both neutrinos and anti-neutrinos. For such a strategy, it is therefore mandatory to discriminate the $\mu^{-}$ and $\mu^{+}$ events at 
the detector, as in the Neutrino Factory proposal. The Magnetised Iron Neutrino Detectors (MIND) studied for use with
Neutrino Factories have thresholds $>$ 3 GeV~\cite{TASD}. Neutrino energies set to first oscillation maximum will 
be below the MIND threshold  for baselines $L<1500$ km. Magnetised
liquid argon detectors and totally 
active scintillator detectors \cite{TASD} have been put forward as alternatives and could provide the techniques to deal 
with this issue. However, with only $\sim 1~\%$ of the beam mono-energetic anti-neutrinos and the possibility of running electron capture and BBD ions
separately, this is a mute point.

The shortest long baseline being considered for the future long baseline neutrino oscillation program is CERN-Frejus at 130 km. 
Using the current values of the oscillation
parameters~\cite{current}, the energy of first oscillation maximum for the $\bar{\nu}_{e}\rightarrow\bar{\nu}_{\mu}$
channel at 130 km is 0.25 GeV. With a boost $\gamma=400$, $Q_{B}>0.315$ MeV is necessary 
for the mono-energetic anti-neutrino flux to get placed on first oscillation maximum at Frejus. For Z=90, BBD will make up $\sim 75$\% of the anti-neutrino flux. Therefore 
the
CBD fraction will be at least 25~\% for all cases in which the mono-energetic neutrinos are to be placed on first maximum.
From Tab.~\ref{T:ions}, all the ions identified have much larger Q-values. The minimum CBD fraction, from these
choices is a much higher $85\%$. All these ions could place a mono-energetic source on (or around) the first oscillation
maximum for the CERN-Canfranc baseline ($L=650$ km).
The CERN-Boulby baseline ($L=1050$ km) requires a minimum $Q_{C}\sim 2.5$ MeV. From the ions selected, no more 
than 1 \% BBD would be possible in this case.
Therefore, in all conceivable cases, a substantial flux from the CBD is to be expected. If these are not separated then one is not exploiting the the 
mono-energetic nature of BBD neutinos.

In the rest frame, the two channels are split by the difference between the CBD and the BBD Q-values, the electron binding energy
$\vert B_{1;Y}\vert$.  Therefore, for a detector with energy resolution $\Delta E$, to separate the channels one requires 
\begin{equation}
\Delta E < 2\gamma \vert B_{1;Y}\vert~.
\end{equation}
For example, $^{207}$Tl$^{81+}$ has $\vert B_{1:Y}\vert=99$ keV. For a detector with $\Delta E = 150$ MeV, a boost $\gamma >
750$ is required. Since $\vert B_{1;Y}\vert\propto (Z+1)^{2}$, where $Z$ is the proton number of the parent, the $\gamma$ factors 
required
will be larger than this for other ions. With the accelerators expected to be available to the community, such as an 
upgraded SPS and the Tevatron, CBD and BBD cannot be separated for this example. A substantial portion of the
anti-neutrino flux will always, therefore, be sourced from the CBD. If 
creating hydrogen like ions is problematic, the BBD neutrinos will be suppressed, or effectively reduced to nil. In that 
case, one would have a high Z anti-neutrino beta beam.    


\section{Conclusions}\label{S:Conclusion}

In the present article, the possibility of using bound beta decays as a source of mono-energetic anti-neutrinos for a 
future long baseline neutrino oscillation experiment has been qualitatively analysed and a number of problem diagnosed. 
The bound beta decay process has been identified as a possible source of mono-energetic anti-neutrinos for a 
CP-even beam~\cite{CPeven}.  The required fluxes will be very hard
to achieve, the low branching ratios and space charge restrictions mean that a target of $10^{18}$ useful decays per year (the 
standard minimum for beta beam related studies) will need major R\&D work. 
Large duty factors will need to be accommodated which will increase the background component of the 
event rate. For anti-neutrinos and neutrinos in the same beam, either sourced from the same ion or for two ion species
circulating simultaneously, discrimination of $\mu^{-}$ and $\mu^{+}$ events is mandatory. For the energies considered, 
this will require innovative technologies such as magnetised totally active scintillator detector or magnitised liquid argon detectors.
For ions that only beta decay and BBD, the two channels need to separated otherwise the mono-energetic nature of the BBD is not being exploited. For the
largest binding energies and a energy resolution of $\Delta E =150$ MeV, this requires a boost $\gamma> 750$. The LHC will be necessary and the event rate will be 
further diminished by the $1/\gamma$ dependence of the useful decay rate. In short, the principal reasons why a bound beta beam is not feasible is the inability
to achieve a useful decay rate for the BBD channel, and the restrictive requirements on the energy resolution of the far detector.


\section*{Acknowledgements}

This work was carried out under the funding of a STFC studentship. The author would like to thank Mats Lindroos, Sergio
Palomares-Ruiz and Silvia Pascoli for their comments.  I would like to further my thanks to Sergio Palomares-Ruiz for 
initially drawing my attention to the suggestion of using bound beta decay as a source for a long baseline neutrino 
oscillation experiment. I acknowledge the support of BENE, CARE contract number RII3-CT-2003-0506395, and the hospitality 
provided by the ISOLDE group at CERN in the latter stages of this work.

\section*{Note added:}

\noindent During the final stages of the present work, a revised and more detailed version of~\cite{CPeven} appeared.


\end{document}